\documentclass[11pt]{article}
\usepackage{a4wide}
\usepackage{amsmath, amssymb, amsfonts, amsthm, amscd, multicol}
\usepackage{diagxy}

\newcommand{\Rt}{\mathcal R^t}
\newcommand{\Z}{\mathcal Z}
\newcommand{\In}[4]{#1^{(#2,#3)}_{#4}}
\newcommand{\R}[4]{\mathfrak{#1}^{(#2,#3)}_{#4}}

\begin{document}

\title{The Algebra of Observables of the Closed Bosonic Nambu-Goto-String: The Presentation of the Algebra of Classical Observables for Light-Like Energy-Momentum in $3+1$ Dimensions\footnote{This work is based on the author's diploma thesis \cite{diplom}}}
\author{Imke Schneider\\Physikalisches Institut, Albert-Ludwigs-Universit\"at,\\Hermann-Herder-Str.~3, D-79104 Freiburg i.~Br., Germany}
\maketitle

\begin{abstract} The presentation of the algebra of classical observables of the closed bosonic Nambu-Goto-String in $3+1$ dimensions is given for the massless case $P^2=0,P\neq0$ in the relevant standard reference frame up to and including the second stratum. The elements and the relations of this algebra established so far are presented in the form of irreducible multiplets of the stabilizer group $E_2$ of the light-like vector $P$ in its standard form.
\end{abstract}
\section{Introduction}We consider the closed bosonic classical relativistic string as a one-dimensional extended \linebreak object moving in $3+1$-dimensional Minkowski space-time with metric $\eta^{\mu\nu}$ of signature\linebreak $(1,-1,-1,-1)$.
  It has been shown in  \cite{pohl.group},\cite{pohl.inv} that the string possesses infinitely many, re\-para\-metri\-za\-tion invariant charges  $\Z_{\mu_1\dots\mu_N}^{(K)}$, which provide a set of  infinitesimal generators of active symmetry transformations. Hence they are also referred to as classical observables. Here $N$ is the rank of the tensor while $K$ denotes the degree of homogeneity. 
 These invariant charges form an associative algebra under tensor multiplication and a graded Poisson algebra $h^\pm_P$ (for given momentum $P$) under taking Poisson brackets.  The rightmover part $h^-_P$ and the leftmover part $h^+_P$ of the algebra commute and their structure constants only differ by sign. Therefore it is sufficient to analyze just the rightmover part being denoted by $h_P$ in the following. The only invariants of degree $K=1$ are the components $P_\mu=\Z_\mu^{(1)}$ of the energy-momentum which are algebraically independent  of all elements of the algebra. Moreover, they Poisson commute with all invariants. Thus we can consider them in the following as numbers and not as elements of the algebra. Doing so, the grading of the algebra is given, both with respect to multiplication and with respect to  Poisson bracket operation, by a single degree $l:=N-K-1$, explicitly
\label{grad}
\begin{eqnarray*}
h_P=  \bigoplus\limits_{l = 0}^{\infty} \mathfrak{V}^{(l)}(h_P), 
\end{eqnarray*}
\begin{eqnarray*}
\{ \mathfrak{V}^{(l)}(h_P), \mathfrak{V}^{(l^\prime)}(h_P)\}\;\subset\;\mathfrak{V}^{(l+l^\prime)}(h_P),\hspace{0.5cm}
  \mathfrak{V}^{(l)}(h_P)\cdot  \mathfrak{V}^{(l^\prime)}(h_P)\;\subset\;  \mathfrak{V}^{(l+l^\prime+1)}(h_P),\hspace{0.5cm}\forall l\geq0.
\end{eqnarray*}
Here $\mathfrak{V}^{(l)}(h_P)$ denotes the vector space of all invariant tensors $\Z_{\mu_1\dots\mu_N}^{(K)}$  with $N>2,K\leq N-1$. All these vector spaces are finite dimensional and referred to as strata.  The vector space $\mathfrak{V}^{(0)}(h_P)$ is $3$-dimensional and forms a subalgebra isomorphic to the Lie algebra of the stabilizer group of $P_\mu$.

By a proposition for the massive string a set of 'standard' invariants exists forming an infinite system of algebraically independent invariants which with respect to multiplication freely generate the algebra of invariants \cite{pohl.alg}. Additionally, for the massive string the presentation of the algebra has been given, i.e.~a set of invariants generating the algebra with respect to Poisson bracket operation has been identified and all relations of degree $l\leq6$ between the generated elements have been computed \cite{pohl.quant}. In the present paper the algebra of classical observables is analyzed for the string with light-like energy-momentum. We construct a set of standard invariants analogously to the massive case and we give the presentation of the algebra in the standard reference frame of light-like $P$ up to and including the second stratum.  The generating elements and relations of the algebra, established so far, are arranged in irreducible multiplets of $E_2$, the stabilizer group of light-like energy-momentum.

\section{An Algebraic Basis for the Algebra of Invariant Charges for  the Case $P^2=0$}
The analysis of the algebra of invariant charges can be considerably facilitated by choosing a special reference frame, i.e.~for the massive string one passes to the rest frame where $P_\mu=\delta_{\mu,0}\,m,\;m>0$. Since the invariant tensors $\Z_{\mu_1\dots\mu_N}^{(K)}$ transform covariantly under the orthochronous Poincar\'e group and the elements $P_\mu$ are central in  the algebra $h_P$, one may  without loss of generality specialize to the momentum rest frame. In \cite{pohl.alg} a minimal and complete algebraic basis for all conserved charges  has been constructed for the massive string. Here we start the analysis for the  massless string with  $P_\mu P^\mu=0,P\neq0$. We exclude the case $P_\mu P^\mu=0,P=0$ being here of no further interest. In the case of light-like energy-momentum no rest frame exists. Instead, the energy-momentum can be chosen to be $P_\mu=\frac{1}{\sqrt{2}}(1,0,0,1)$. This can be done without loss of generality for the above mentioned reasons. Note that an arbitrary light-like vector can always be transformed into this form by a suitable Lorentz transformation. 
First we introduce a new basis in Minkowski space by the basis transformations $e_0\rightarrow e_{\circ}:= \frac{1}{\sqrt{2}}(e_0+e_3)$ and $e_3\rightarrow e_\times:=\frac{1}{\sqrt{2}}(e_0-e_3)$, so that the standard light-like vector $P$ assumes the form $(1,0,0,0)$ in the new coordinates. This makes it possible to construct  an algebraic basis for all conserved charges for the massless string in complete analogy to the massive case. Employing additionally the angular momentum basis vectors  $e_{-1}:=\frac{1}{\sqrt{2}}(e_1-ie_2)$ and $e_{+1}:=-\frac{1}{\sqrt{2}}(e_1+ie_2)$, which  will facilitate the presentation of the algebra later on, we obtain the metric coefficients in the new basis  $\{e_\circ,e_{-1},e_{+1},e_\times\}$ as follows:
\begin{eqnarray} \eta_{\mu\nu}=\left(\begin{array}{cccc} 0&0&0&1\\ 0&0&1&0\\
0&1&0&0\\ 1&0&0&0\\ \end{array}\right)                   \hspace{1cm}\mu,\nu
\in \{\circ, -, + ,\times\}.
 \end{eqnarray}
 Note that $e_\circ$ and $e_\times$ are conjugate to each other, as are $e_{+1}$ and $e_{-1}$.

To continue the construction of a minimal and complete algebraic basis, we follow the course of construction which has been taken in the massive case and replace the subscripts by the following mappings $0\rightarrow\circ,1\rightarrow-,2\rightarrow+,3\rightarrow\times$.   
As in \cite {pohl.alg} the homogeneous invariants $\Z^{(K)}_{\mu_1\dots\,\mu_N}$ are given by polynomials of degree $K$ of the so called truncated tensors $ \Rt_{\mu_1\dots\,\mu_N}$:
\begin{eqnarray}\label{expansion}
\Z^{(K)}_{\mu_1\dots\,\mu_N}=Z_N\circ\left(\frac{1}{K!}
                              \sum_{0<a_1<\dots\,<a_{k-1}<N}
                              \Rt_{\mu_1\dots\,\mu_{a_1}}
                              \Rt_{\mu_{a_1+1}\dots\,\mu_{a_2}}\dots
                              \Rt_{\mu_{a_{k-1}+1}\dots\,\mu_N}\right),
\end{eqnarray}
where $Z_N$ denotes the cyclic sum over $N$ indices.
The truncated tensors are distinguished by the fact that there are no other dependencies than linear ones among them. Hence, by choosing a basis for the truncated tensors, it is possible to express the invariant charges $\Z^{(K)}_{\mu_1\dots\,\mu_N}$ explicitly as polynomials in algebraically independent tensors. The truncated tensors themselves, however, are not reparametrization invariant.
With
\begin{alignat*}{7}
  &\mathcal{P}_i &\quad = \quad& \Z^{(1)}_i &\quad = \quad& \Rt_i&\quad = \quad&0 \qquad i\in\{-,+,\times\} \\ 
  &\mathcal P_\circ &\quad = \quad& \Z^{(1)}_\circ &\quad = \quad& \Rt_\circ &\quad = \quad& 1 
\end{alignat*}
all terms containing a factor $\Rt_i,\,i\in\{-,+,\times\}$ disappear in (\ref{expansion}). The polynomial degree of terms containing a factor $\Rt_\circ$ decreases by one. Thus, the previously homogeneous polynomial becomes an inhomogeneous polynomial. Every invariant can now be assigned to a so called \emph{dominant part} consisting of the terms with minimal polynomial degree, i.e.~of a sum of \emph{dominant monomials}. By means of this assignment, a system of algebraically independent invariants can be specified which generates the  algebra of invariants via  multiplication.
 First a maximal basis of those tensors $\Rt_{\mu_1\dots\,\mu_{K+1}}$ with  $\mu_1,\mu_{K+1}\in\{-,+,\times\}$ is chosen. Each of these tensor components $\Rt_{K+1,i}\;(i=1,\dots\, )$ is assigned to a so called standard invariant\footnote{Latin letters
  $i,\, j$ are used, if $i,j\in\{-,+,\times\}$ analogously to the usual notation of space-like indices. }:
\begin{eqnarray}
\Rt_{i\mu_2\dots\,\mu_{K}j} \rightarrow \left\{
\begin{array}{ll}\frac{1}{K}\Z^{(2)}_{\circ i\mu_2\dots\,\mu_{K}j}& \mbox{for
} K=1 \mbox{ or } \mu_2=\dots\,=\mu_{K}=\circ \\ &   \\
(K-1)!\;\Z^{(K)}_{\underbrace{\circ\dots\circ}_{(K-1)\times}i\mu_2\dots\,\mu_{K}j}&
\mbox{for } K>1 \mbox{ and }
(\mu_2,\dots,\mu_{K})\ne(\circ,\dots,\circ).\end{array}         \right.
\end{eqnarray}
In any case, the dominant part of the standard invariants only consists of a single monomial, which is exactly the corresponding tensor component $\Rt_{i\mu_2\dots\,\mu_{K}j}$ of the basis. According to the algebraical independence of the tensor components $\Rt_{K+1,i}\;(i=1,\dots\, )$, the standard invariants are algebraically independent.
By proposition $17$ of \cite{pohl.alg} for the massive case, meanwhile rigorously proved \cite{pohl.priv}, any invariant polynomial in the truncated tensors $\Rt_{\mu_1\dots\,\mu_{N}}$ can be uniquely expressed as a polynomial in the standard invariants. For the massless case this proposition applies too: employing the basis $\{e_\circ,e_{-1},e_{+1},e_\times\}$ the metric changes but this is irrelevant for the application of the proposition. Thus the standard invariants provide an algebraical basis of the algebra of invariants in the standard reference frame.

Now we can determine the number $n_l$ of independent standard invariants of any stratum $\mathfrak{V}^{(l)}(h_P)$ in the massless case, independent in particular of the invariant charges of the strata with lower degree. This number coincides with the number of linear independent  tensors $\Rt_{\mu_1\dots\,\mu_{l+1}}$ with indices $\mu_1,\mu_{l+1}\in\{-,+,\times\}$ at the extremal positions. The number $n_l$ is equal to the number of standard invariants of the same stratum in the massive case. The latter has been computed  in \cite{pohl.alg} for general space-time dimension $d$ and is given by:
\begin{eqnarray}\label{anzst}
n_l(d)=n(d,l+2)-n(d,l+1),\mbox{ where}
  &n(d,N):=\frac{1}{N}\sum\limits_{D|N}\mu(D)d^{\frac{N}{D}}. 
\end{eqnarray}
Here the sum extends over all divisors $D$ of $N$, and $\mu(D)$ denotes the M\"obius function:
\begin{eqnarray*}
\mu(D)=
  \begin{cases} 
    1 & \textrm{for $D = 1$}\\
    (-1)^n & \textrm{if $D$ can be decomposed in exactly $n$ different prime factors}\\
    0 & \textrm{if at least two prime factors of $D$ coincide.}\\
  \end{cases}
\end{eqnarray*}
For example one obtains $n_0(4)=3,\;n_1(4)=14$ and $n_2(4)=40$.  


\section{The Action of the Little Group $E_2$ on the Standard Invariants for $P^2=0$ }

The algebra of standard invariants carries a representation of the orthochronous Poincar\'e group. It must be noted that for the massive case the energy-momentum vector $P_\mu=\delta_{\mu,0}m$ remains stable under the parity transformation, i.e.~the $3$-dimensional space reflection, while for the massless case the energy-momentum vector $P_\mu=\delta_{\mu,\circ},\mu\in \{\circ, -, + ,\times\}$ is not invariant under the parity transformation. Instead, it remains stable under the parity transformation combined with a suitable orthochronous Lorentz transformation. We will account for these transformations by including reflections at the $1$-axis of the $2$-dimensional Euclidean plane in the stabilizer group of the pertinent energy-momentum vector.
For the case $P_\mu P^\mu=0,P\ne0$ the little group of the orthochronous Poincar\'e group is the $2$-dimensional Euclidean group $E_2$. Using the convention of \cite{edm} we understand by $E_2$ the group of translations and rotations in two dimensions, respectively, including reflections at the $1$-axis of the $2$-dimensional Euclidean plane. The infinitesimal generators of the little groups of the orthochronous Poincar\'e group are obtained by the components of the Pauli-Lubanski-Vector 
\begin{eqnarray}\label{pa-lu} 
W^{\mu}=\frac{1}{2}{\epsilon}^{\mu\nu\rho\sigma}P_{\nu}M_{\rho\sigma}
\hspace{1cm} \mu,\nu,\rho,\sigma\in\{0,1,2,3\}.
\end{eqnarray} 
With $P_\mu=\frac{1}{\sqrt{2}}(1,0,0,1)$ in Cartesian coordinates, the three linear independent components $W_1,W_2$, generating the translations, and $J_3:=\frac{1}{\sqrt{2}}(W_0+W_3)$, generating the rotations, provide a basis of the Lie algebra $\mathfrak{e}_2$. The conventional basis chosen for the $\mathfrak{e}_2$ is given by :
\begin{eqnarray}
\{J_3,\,W_{-1}:=\frac{1}{\sqrt{2}}(W_1-iW_2),\,W_{+1}:=-\frac{1}{\sqrt{2}}(W_1+iW_2)\}.
\end{eqnarray}
 Accordingly, the following commutation rules are obtained:
\begin{eqnarray}\label{alg}
i\{W_{+1},W_{-1}\}=0,\quad
i\{J_3,W_{+1}\}=W_{+1},\quad 
i\{J_3,W_{-1}\}=-W_{-1}.
\end{eqnarray}
The Lie-Algebra $\mathfrak{e}_2$ may be considered as a contraction of the Lie-Algebra $\mathfrak so(3)$ in the limit  $m\rightarrow 0$. Unlike $\mathfrak so(3)$, it is neither semi-simple nor compact.

In the following, we examine how the generators of  $\mathfrak{e}_2$ operate on the standard invariants by Poisson bracket operation.
In terms of the standard invariants the components of the Pauli-Lubanski-Vector are invariants of degree $l=0$, namely
\begin{eqnarray}
W^{\mu} = -\frac{1}{12}{\epsilon}^{\mu\nu\rho\sigma}
                \Z^{+(2)}_{\nu\rho\sigma}
\hspace{1cm}\mu,\nu,\rho,\sigma\in\{0,1,2,3\}.
\end{eqnarray}
 For the case considered here it follows in the selected basis that
\begin{eqnarray}
  W_{+1}=-\frac{i}{2}\Z^{(2)}_{\circ+\times},\hspace{0,5cm}
  W_{-1}=\frac{i}{2}\Z^{(2)}_{\circ-\times},\hspace{0,5cm}
  J_3= -\frac{i}{2}\Z^{(2)}_{\circ-+}.
\end{eqnarray}
Complex conjugation yields a $*$-operation on the algebra which includes the reflections at the $1$-axis.
First the invariants $\Z^{(K)}_{\mu_1\dots\mu_N}$ of any stratum are organized in tensor operators $O_m$ with
\[i\{J_3,O_m\}=m\;O_m.\]
Employing the angular momentum coordinates $e_{+1}$ and $e_{-1}$ the basis of the Minkowski space is chosen to be such that the standard invariants are eigenvectors of $J_3$ corresponding to the eigenvalue $m$:
\begin{eqnarray}
i\{J_3,\Z^{(K)}_{\mu_1\dots\mu_N}\} &=&m \;\Z^{(K)}_{\mu_1\dots\mu_N.}\\
&=&\sum_{i=1}^N(\delta_{+,\mu_i}-\delta_{-,\mu_i})\;\Z^{(K)}_{\mu_1\dots\mu_N}.
\end{eqnarray}
 The generators $W_{+1}$ and $W_{-1}$ operate by Poisson bracket to some extent like raising  and lowering operators. Their action on a tensor operator $O_m$ leads to a linear combination of eigenvectors corresponding to eigenvalue $(m+1)$ and $(m-1)$, respectively, unless $i\{W_{\pm1},O_m\}=0$. Similarly, the Poisson bracket $i\{O_m,O_{m^\prime}\}$ belongs to the eigenspace $m+m^\prime$ or equals $0$. Normalization of the tensor operators $O_m$ is such that
\begin{eqnarray}
O_m^*=(-1)^mO_{-m}.
\end{eqnarray}
But it should be pointed out that unlike the raising and lowering operators of  $\mathfrak so(3)$ the action of $W_{+1}$ and $W_{-1}$ only works one-way. Once one of the two generators is applied to any tensor $O_m$ it is impossible to reobtain this tensor by the action of the other generator. 

It is well known that the irreducible representations of the proper orthochronous Poincar\'e group $P^\uparrow_+$ can be reduced to the irreducible representations of the little groups according to the orbits $(P^2>0,P_0\gtrless0),(P^2=0,P_0\gtrless0),P^2<0$ and $P=0$, respectively. A classification of all unitary representations of the Lorentz group has been carried out by Wigner  \cite{wigner}. Among the massless representations there are two classes. One is characterized by $P^2=0,W^2=0$, where the generators of the translation are represented trivially. The other one, characterized by $P^2=0,W^2=\rho^2,\rho>0$, describes particles of zero rest mass and continuous spin. One might expect that the algebra of invariants carries a representation of the first type, but this is not the case:~the generators $W_{+1}$ and $W_{-1}$ do not operate trivially on the standard invariants. Since the standard invariants provide an algebraic basis of the algebra, this can be easily verified by some simple examples. It is sufficient to obtain the action on the leading terms, the dominant monomials. The pertinent computations are performed with the help of the   modified Poisson bracket $\{\;,\;\}_*$, introduced in \cite{pohl.alg}, for the truncated tensors $\Rt_{\mu_1\dots\mu_N}$:
\begin{eqnarray*}
m=2:\hspace{2cm} i\{W_{+1},\Rt_{++\times}\}_*&=&0\\
                i\{W_{-1},\Rt_{++\times}\}_*&=&-\Rt_{+\times\times}\\
\\
               i\{W_{+1},\Rt_{+\circ+}\}_*&=&0\\
                i\{W_{-1},\Rt_{+\circ+}\}_*&=&2\Rt_{++-}+2\Rt_{+\circ\times}\\ 
\\
m=1:\hspace{2cm} i\{W_{+1},\Rt_{+\circ\times}\}_*&=&-\Rt_{++\times}\\
        i\{W_{-1},\Rt_{+\circ\times}\}_*&=&-\Rt_{\times\circ\times}+\Rt_{+-\times}.
\end{eqnarray*}
The vector space basis of $\mathfrak{V}^{(1)}(h_P)$ can be ordered in subspaces which are left invariant  by the action of  $E_2$. An inconvenient feature of $E_2$ is the fact that there is no guarantee for the decomposability of a given representation of $E_2$. Unitary representations of any group are always fully reducible, but unitarity does not have to be demanded for the algebra of observables,  it has to be required only for physical states. 


\section{The Presentation of the Algebra of  Standard Invariants for $P^2=0$}
In terms of the standard invariants  a system of algebraically independent invariants is known which freely, but not finitely, generate the algebra of invariant charges via multiplication. Since the algebra of invariant charges is a graded Poisson algebra, the question occurs, to what extent invariants of higher degree can be expressed in terms of  Poisson brackets of invariants of lower degree, particularly  in terms of invariants of the zeroth and first stratum. 
In the (classical) massive case it has been shown in \cite{pohl.alg} that the closure under forming multiple Poisson brackets of the elements of the Lorentz-group representation-spaces of the zeroth  and first stratum does not supply an algebraic basis of the entire set $h_P$. In every stratum of odd degree there exists at least one exceptional element. Moreover, in general  a Poisson bracket of two standard invariants does not only  yield a linear combination of standard invariants but also a linear combination of  products of standard invariants of lower degree. Thus, to generate the algebra, one needs (multiple) Poisson brackets as well as products of a set of generating elements. Here we show for the massless case how the second stratum can be produced by invariant charges of lower degrees. Additionally,  we give a presentation of the algebra of invariants up to the second stratum in terms of relations imposed on invariants of the second stratum generated by the standard invariants of the zeroth and first stratum.  At the same time $\mathfrak{V}^{(1)}(h_P)$ and $\mathfrak{V}^{(2)}(h_P)$ are arranged in multiplets of $\mathfrak{e}_2$. 

 The vector space $\mathfrak{V}^{(0)}(h_P)$ coincides with the $\mathfrak{e}_2$ generating elements $J_3$, $W_{-1}$ and $W_{+1}$. The vector space $\mathfrak{V}^{(1)}(h_P)$ is spanned by  $14$ standard invariants of degree $l=1$ and, additionally, by the products $J_3^2,W_{-1}^2,W_{+1}^2,J_3 W_{-1},J_3 W_{+1}\mbox{ and }W_{-1}W_{+1}$. In the sequel an explicit though somewhat arbitrary basis of these $14$ standard invariants is constructed under the aspect of generating the second stratum as simple as possible. An unpleasant feature of the massless case is the fact that the linear span of the standard invariants of the first stratum does not provide a representation space of $E_2$ on its own, but only the vector space $\mathfrak{V}^{(1)}(h_P)$ as a whole provides such a representation space.  Moreover, it is impossible to avoid that the $\mathfrak{e}_2$ invariant subspaces of the first stratum have elements in common. These features will be explained in more detail later on. The particular choice of the basis of the standard invariants of $\mathfrak{V}^{(1)}(h_P)$ is done according to the criterion that the basis vectors are arranged in preferably maximal $\mathfrak{e}_2$ invariant subspaces which overlap as little as possible as far as the standard invariants are concerned. Speaking in terms of matrices, $W_{+1}$ and $W_{-1}$ are approximately in block form with the matrix elements being predominately equal to $1$ or $0$, while $J_3$ stays a diagonal matrix. This choice has the advantage that the standard invariants of the second stratum being generated by Poisson bracket operation are automatically arranged according to the same criteria. This facilitates the task of finding the relations, which do not emerge by the action of $\mathfrak{e}_2$ from one another, in particular the generating ones. 

 The $20$ elements of the vector space basis of $\mathfrak{V}^{(1)}(h_P)$  are arranged in multiplets of $E_2$, which are listed in the sequel. The more interesting basis vectors which provide a basis for the standard invariants of the first stratum are written bold faced. Further notation is explained below. The multiplets are as follows: \\
A $9$-dimensional multiplet $T$:
\begin{alignat*}{1}
  T\quad=\quad&\{{\bf \In{T}{0}{0}{0},\In{T}{1}{0}{1},
  \In{T}{0}{1}{-1},
  \In{T}{1}{1}{0},\In{T}{2}{0}{2},\In{T}{0}{2}{-2},\In{T}{2}{1}{1},\In{T}{1}{2}{-1}},W_{-1} W_{+1}\}\\
  &\mbox{with }\In{T}{0}{0}{0}:=\frac{1}{2}\Z^{(2)}_{\circ-\circ+}.\\ 
\end{alignat*}
A $10$-dimensional multiplet $(A,\bar{A})$ consisting of two sets $A$ and $\bar{A}$ with $\bar{A}=A^*$:
\begin{alignat*}{2}
  A\quad=\quad&\{{\bf \In{A}{0}{0}{2},\In{A}{0}{1}{1},\In{A}{0}{2}{0}},
  -6\In{T}{1}{2}{-1}-96J_3 W_{-1},W_{-1}^2\}\\
  &\textrm{with }\In{A}{0}{0}{2}:=\frac{1}{2}\Z^{(2)}_{\circ+\circ+},\\
  \\
  \bar{A}\quad=\quad&\{{\bf\In{\bar{A}}{0}{0}{-2},\In{\bar{A}}{1}{0}{-1},\In{\bar{A}}{2}{0}{0}},
  -6\In{T}{2}{1}{1}+96J_3 W_{+1},W_{+1}^2\}\\
  &\textrm{with }\In{\bar{A}}{0}{0}{-2}:=\frac{1}{2}\Z^{(2)}_{\circ-\circ-}.\\
\end{alignat*}
A $6$-dimensional multiplet $U$, containing only products of the elements of $\mathfrak{V}^{(0)}(h_P)$ :
\[U=\{J_3^2,-2J_3 W_{+1},2J_3 W_{-1},2W_{+1}^2,2W_{-1}^2,-2W_{-1}W_{+1}\}.\]
The notation of the invariants is as follows: the lower index denotes the eigenvalue of $J_3$, and the two upper indices indicate the multiplicity of repeated action of $W_{+1}$ (first upper index) and
$W_{-1}$ (second upper index) on the invariant charges generating the multiplets. Since $W_{+1}$ and $W_{-1}$ commute, this notation is consistent. If two sets emerge from one another by the $*$-operation, one of them is labeled with a barred letter. The multiplet  $(A,\bar{A})$ contains the elements $-6\In{T}{1}{2}{-1}-96J_3 W_{-1}$ and $-6\In{T}{2}{1}{1}+96J_3 W_{+1}$ which differ only by products of $W_{+1},W_{-1}$ and $J_3$ from elements of the multiplet $T$. We call such a situation an \emph{improper overlap}.  Likewise, we call correspondingly generalized  situations for higher strata an improper overlap, i.e.~situations, in which multiplets contain elements which can be expressed as sums of standard invariants of the same degree of other multiplets and of products of elements of lower degree multiplets.   Such an improper overlap, however, cannot be avoided by choice of a different basis. Here we see that the linear span of the standard invariants of the first stratum does not provide a representation space of $E_2$ on its own, but that products of $W_{+1}$, $W_{-1}$ and $J_3$ are needed to complete the $E_2$ invariant subspaces. The more interesting multiplets $T$ and $(A,\bar{A})$ provide $19$ basis vectors, so the multiplet $U$, obtained by repeated action of $W_{+1}$ and $W_{-1}$ on $J_3^2$ is needed to complete the $20$-dimensional vector space basis of $\mathfrak{V}^{(1}(h_P)$. If we consider all three multiplets, we see that the multiplets (properly) overlap.

Computed explicitly, the basis elements are given by: 
\begin{multicols}{2}
\parbox{7cm}{
\begin{eqnarray*}
  \In{T}{0}{0}{0}&=&\frac{1}{2}\Z^{(2)}_{\circ-\circ+}\\
  \In{T}{1}{0}{1}&=& - \Z^{(2)}_{\circ-++} + \frac{1}{2}\Z^{(2)}_{\circ+\circ\times} \\
  \In{T}{0}{1}{-1}&=& \Z^{(2)}_{\circ- - +} - \frac{1}{2}\Z^{(2)}_{\circ-\circ\times} \\
  \In{T}{2}{0}{2}&=& -2 \Z^{(2)}_{\circ+ + \times}  
\end{eqnarray*}
}
\parbox{7cm}{
\begin{eqnarray*}
  \In{T}{1}{1}{0}&=& - \frac{1}{2}\Z^{(2)}_{\circ\times\circ\times} -4J^2_3\\
  \In{T}{0}{2}{-2}&=& -2 \Z^{(2)}_{\circ- - \times} \\ 
  \In{T}{2}{1}{1}&=& - 2\Z^{(2)}_{\circ+\times\times} +8J_3 W_{+1}\\
  \In{T}{1}{2}{-1}&=& 2 \Z^{(2)}_{\circ-\times\times}-8J_3 W_{-1}
\end{eqnarray*}
}
\end{multicols}
\begin{multicols}{2}
\parbox{7cm}{
\begin{eqnarray*}
  \In{A}{0}{0}{2}&=&\frac{1}{2}\Z^{(2)}_{\circ+\circ+}\\
  \In{A}{0}{1}{1}&=&-2\Z^{(2)}_{\circ++-}-\Z^{(2)}_{\circ+\circ\times}\\
  \In{A}{0}{2}{0}&=&-4\Z^{(2)}_{\circ+-\times}+\Z^{(2)}_{\circ\times\circ\times} -8J^2_3
\end{eqnarray*}}
\parbox{7cm}{
\begin{eqnarray*}
  \In{\bar{A}}{0}{0}{-2}&=&\frac{1}{2}\Z^{(2)}_{\circ-\circ-}\\
  \In{\bar{A}}{1}{0}{-1}&=&2\Z^{(2)}_{\circ+--}+\Z^{(2)}_{\circ-\circ\times}\\
  \In{\bar{A}}{2}{0}{0}&=&-4\Z^{(2)}_{\circ-+\times}+
  \Z^{(2)}_{\circ\times\circ\times}-8J^2_3.
\end{eqnarray*}}
\end{multicols}
In the following diagrams the structure of the selected basis is illustrated.
\paragraph{1.~The multiplet $\boldsymbol{T}$:\label{basis1}} 

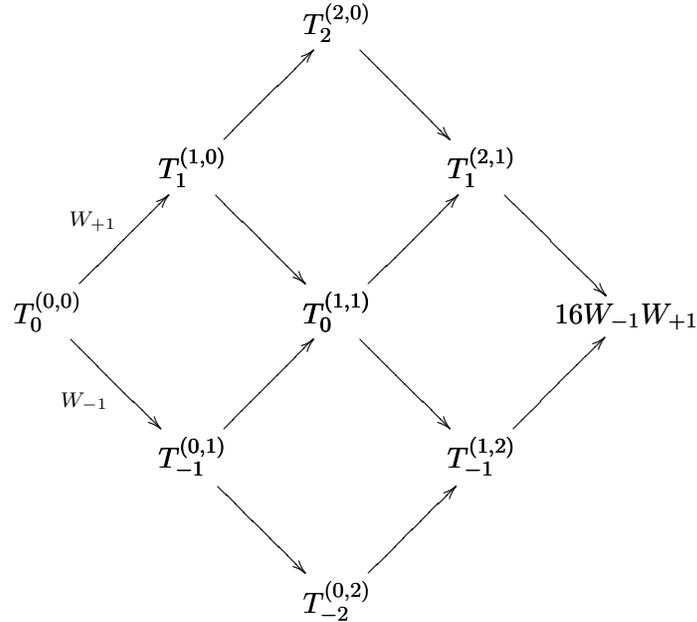
$$\bfig
\morphism<500,500>[\In{T}{0}{0}{0}`\In{T}{1}{0}{1};W_{+1}]
\morphism|b|<500,-500>[\In{T}{0}{0}{0}`\In{T}{0}{1}{-1};W_{-1}]
\morphism(500,500)|u|<500,-500>[\In{T}{1}{0}{1}`\In{T}{1}{1}{0};W_{-1}]
\morphism(500,-500)|u|<500,500>[\In{T}{0}{1}{-1}`\In{T}{1}{1}{0};W_{+1}]
\morphism(500,500)|u|<500,500>[\In{T}{1}{0}{1}`\In{T}{2}{0}{2};W_{+1}]
\morphism(500,-500)|u|<500,-500>[\In{T}{0}{1}{-1}`\In{T}{0}{2}{-2};W_{-1}]
\morphism(1000,0)|u|<500,500>[\In{T}{1}{1}{0}`\In{T}{2}{1}{1};W_{+1}]
\morphism(1000,0)|u|<500,-500>[\In{T}{1}{1}{0}`\In{T}{1}{2}{-1};W_{-1}]
\morphism(1000,-1000)|u|<500,500>[\In{T}{0}{2}{-2}`\In{T}{1}{2}{-1};W_{+1}]
\morphism(1000,1000)|u|<500,-500>[\In{T}{2}{0}{2}`\In{T}{2}{1}{1};W_{-1}]
\morphism(1500,500)|u|<500,-500>[\In{T}{2}{1}{1}`16W_{-1}W_{+1};W_{-1}]
\morphism(1500,-500)|u|<500,500>[\In{T}{1}{2}{-1}`16W_{-1}W_{+1};W_{-1}]
\efig$$
An arrow pointing upward indicates the action of $W_{+1}$, while an arrow pointing downward indicates the action of $W_{-1}$. A missing arrow indicates a zero result upon application of $W_{+1}$ and $W_{-1}$. 
\paragraph{2.~The multiplet $(\boldsymbol{A},\boldsymbol{\bar{A}})$:}
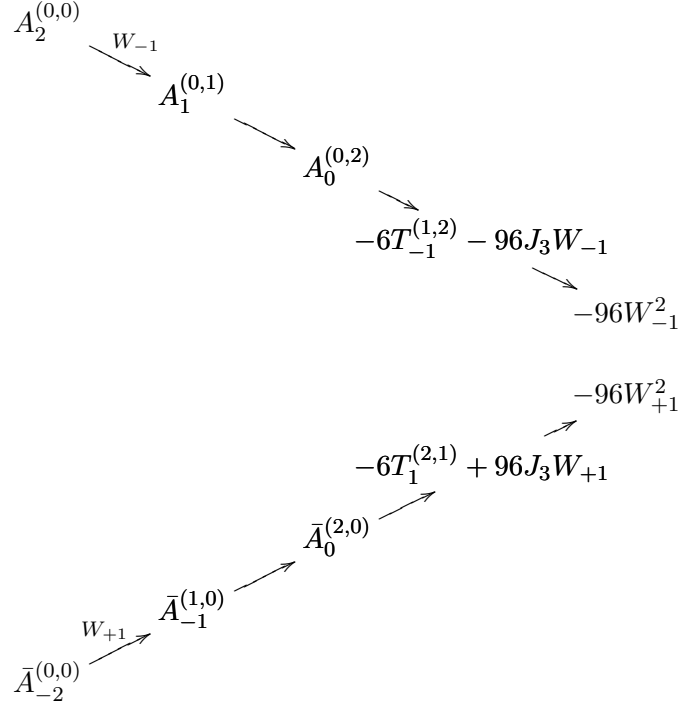
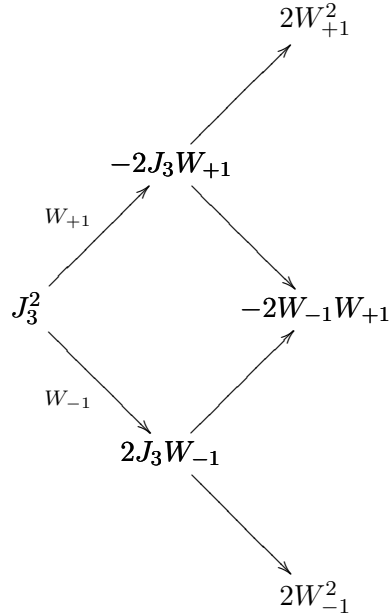
$$\bfig
\morphism<500,-250>[\In{A}{0}{0}{2}`\In{A}{0}{1}{1};W_{-1}]\\
\morphism(500,-250)|u|<500,-250>[\In{A}{0}{1}{1}`\In{A}{0}{2}{0};W_{-1}]\\
\morphism(1000,-500)|u|<500,-250>[\In{A}{0}{2}{0}`-6\In{T}{1}{2}{-1}-96J_3 W_{-1};W_{-1}]\\
\morphism(1500,-750)|u|<500,-250>[-6\In{T}{1}{2}{-1}-96J_3 W_{-1}`-96W_{-1}^2;W_{-1}]
\efig$$
$$\bfig
\morphism<500,250>[\In{\bar{A}}{0}{0}{-2}`\In{\bar{A}}{1}{0}{-1};W_{+1}]\\
\morphism(500,250)|u|<500,250>[\In{\bar{A}}{1}{0}{-1}`\In{\bar{A}}{2}{0}{0};W_{+1}]\\
\morphism(1000,500)|u|<500,250>[\In{\bar{A}}{2}{0}{0}`-6\In{T}{2}{1}{1}+96J_3 W_{+1};W_{+1}]\\
\morphism(1500,750)|u|<500,250>[-6\In{T}{2}{1}{1}+96J_3 W_{+1}`-96W_{+1}^2;W_{+1}]\\ \\
\efig$$
\paragraph{3.~The multiplet $(\boldsymbol{U})$:}
$$\bfig
\morphism<500,500>[J_3^2`-2J_3 W_{+1};W_{+1}]
\morphism|b|<500,-500>[J_3^2`2J_3 W_{-1};W_{-1}]
\morphism(500,500)|u|<500,-500>[-2J_3 W_{+1}`-2W_{-1}W_{+1};W_{-1}]
\morphism(500,-500)|u|<500,500>[2J_3 W_{-1}`-2W_{-1}W_{+1};W_{+1}]
\morphism(500,500)|u|<500,500>[-2J_3 W_{+1}`2W_{+1}^2;W_{+1}]
\morphism(500,-500)|u|<500,-500>[2J_3 W_{-1}`2W_{-1}^2;W_{-1}]
\efig$$
\section{$\mathfrak{V}^{(2)}(h_P)$, the Stratum of Degree 2}
The dimension of the stratum $\mathfrak{V}^{(2)}(h_P)$ is 92, while the number of standard invariants of $\mathfrak{V}^{(2)}(h_P)$  is $n_l=40$. The latter ones  can be generated by the standard invariants of $\mathfrak{V}^{(0)}(h_P)$ and $\mathfrak{V}^{(1)}(h_P)$, whereby by choice of the basis of the first stratum constructed above they are arranged in multiplets of $\mathfrak{e}_2$. Nevertheless the construction of the basis of $\mathfrak{V}^{(2)}(h_P)$ is somewhat arbitrary, so the process is shortly explained while the explicit form of the multiplets and the standard invariants at the top of  the multiplets are given in the appendix.

\begin{enumerate}
\item Starting with the multiplet T, first the Poisson brackets of the invariant $\In{T}{0}{0}{0}$ at the top of the multiplet and the next elements of the multiplet,  $\In{T}{1}{0}{1}$ and $\In{T}{0}{1}{-1}$, are computed:
\[\In{V}{0}{0}{1}:=\{\In{T}{0}{0}{0},\In{T}{1}{0}{1}\}\neq0\hspace{1cm} 
\In{\bar{V}}{0}{0}{-1}:=\{\In{T}{0}{0}{0},\In{T}{0}{1}{-1}\}\neq0.\]
By repeated action of $W_{+1}$ and $W_{-1}$ on $\In{V}{0}{0}{1}$ and $\In{\bar{V}}{0}{0}{-1}$ an invariant subspace $(V,\bar{V})$ is spanned. All possible Poisson brackets of the basis elements of $T$ with itself, which are not already obtained in $(V,\bar{V})$, are arranged in multiplets, separately. Next the Poisson brackets of the elements at the top of the multiplets $T$ and $(A,\bar{A})$ are computed:
\[\mathfrak{A}^{(0,0)}_2 :=\{\In{A}{0}{0}{2},\In{T}{0}{0}{0}\}=0\hspace{1cm}\mathfrak{\bar{A}}^{(0,0)}_{-2}:= \{\In{\bar{A}}{0}{0}{-2},\In{T}{0}{0}{0}\}=0.\]
These Poisson brackets already yield relations. Therefore they are labeled by a different script style. We continue with the following Poisson bracket:
\[
  \In{B}{0}{0}{0}:=
 \{\In{A}{0}{0}{2},\In{\bar{A}}{0}{0}{-2}\}\neq0.\]
$\In{B}{0}{0}{0}$ is the top component of a multiplet $B$ obtained from repeated action of $W_{+1}$ and $W_{-1}$ on $\In{B}{0}{0}{0}$. The further Poisson brackets of elements of $T$ and  $(A,\bar{A})$, unequal to zero, which are not obtained so far, are arranged in multiplets, separately. The full list of multiplets of the standard invariants of  $\mathfrak{V}^{(2)}(h_P)$ is given in the appendix.

\item All possible relations among the terms established so far are computed with the arrangement in multiplets facilitating the work considerably. The relations are listed in the following.
\item A basis of linear independent standard invariants of the second stratum is chosen, see in the appendix.
\end{enumerate}

\section{The Relations}
The number of possible (multiple-) Poisson brackets is greater than the number $n_l$ of linearly independent standard invariants of the target stratum even when the anti-symmetry and the Jacobi identity are taken into account. The $14$ generators of the first stratum lead to $\frac{1}{2}*14*13=91$ Poisson brackets already accounting for anti-symmetry. Since the second stratum is generated by simple Poisson brackets, no dependencies by Jacobi identity arise. With $n_2=40$ and the fact that all standard invariants of the second stratum can be generated by Poisson brackets, $51$ relations must exist. Just as the generating elements of the algebra, the relations can be presented in form of irreducible multiplets of $\mathfrak{e}_2$.  In the second stratum there are $5$ irreducible multiplets of relations. They are obtained by the action of $W_{+1},W_{-1}$ and the $*$-operation on the relations at the top of the multiplets:
\begin{eqnarray*}
\mathfrak{A}^{(0,0)}_2 :\hspace{1cm}0&=&i\{\In{A}{0}{0}{2},\In{T}{0}{0}{0}\}
\\ \\
\mathfrak{B}^{(0,0)}_0:\hspace{1cm}0&=&i\{\In{A}{0}{1}{1},\In{\bar{A}}{1}{0}{-1}\}
                                     -2i\{\In{A}{0}{1}{1},\In{T}{0}{1}{-1}\}
                                   -2i\{\In{T}{1}{0}{1},\In{\bar{A}}{1}{0}{-1}\}\\ 
\phantom{\mathfrak{B}^{(0,0)}_0:}\hspace{1cm}&& -4i\{\In{T}{0}{1}{-1},\In{T}{1}{0}{1}\}
\\ \\
\mathfrak{C}^{(0,0)}_2:\hspace{1cm}0&=&i\{\In{A}{0}{0}{2},\In{\bar{A}}{2}{0}{0}\}
                                      +4i\{\In{A}{0}{1}{1},\In{T}{1}{0}{1}\}
                               -\frac{1}{3}i\{\In{A}{0}{0}{2},\In{A}{0}{2}{0}\}\\                                
                              &&-4i\{\In{T}{0}{0}{0},\In{T}{2}{0}{2}\}
                                      +64J_{3} \In{A}{0}{0}{2} \
\\ \\
\mathfrak{D}^{(0,0)}_0:\hspace{1cm}0&=&+8i\{\In{T}{0}{0}{0},\In{T}{1}{1}{0}\} +4i\{\In{A}{0}{1}{1},
                           \In{T}{0}{1}{-1}\} +4i\{\In{\bar{A}}{1}{0}{-1},\In{T}{1}{0}{1}\}
\\ \\
\mathfrak{E}^{(0,0)}_0:\hspace{1cm}0&=&\frac{1}{4}i\{\In{T}{2}{1}{1},\In{\bar{A}}{1}{0}{-1}\}
                         +\frac{1}{2}i\{\In{T}{1}{1}{0},\In{\bar{A}}{2}{0}{0}\}
                          -\frac{3}{2}i\{\In{T}{0}{1}{-1},\In{T}{2}{1}{1}\}\\
                       &&  -\frac{3}{2}i\{\In{T}{1}{2}{-1},\In{T}{1}{0}{1}\}
                     +\frac{1}{2}i\{\In{A}{0}{2}{0},\In{T}{1}{1}{0}\}
                        +\frac{1}{4}i\{\In{A}{0}{1}{1},\In{T}{1}{2}{-1}\}\\
                        &&-2i\{\In{T}{0}{2}{-2},\In{T}{2}{0}{2}\}
                        +\frac{1}{2}i\{\In{A}{0}{2}{0},\In{\bar{A}}{2}{0}{0}\}
                  +36W_{+1}\In{T}{0}{1}{-1}\\
		  &&-2W_{+1}\In{\bar{A}}{1}{0}{-1}
                       -36W_{-1}\In{T}{1}{0}{1}+2W_{-1}\In{A}{0}{1}{1}
                             -2J_3\In{A}{0}{2}{0}\\
               & &-2J_3\In{\bar{A}}{2}{0}{0}-72J_3\In{T}{1}{1}{0}
               -64J^3_3.\\
\end{eqnarray*}
The notation of the relations is analogous to that of the invariants. However, the set of the above relations does not fully decompose into invariant subspaces of $\mathfrak{e}_2$. In the following, it is shown how the $51$ relations of the $2.$~stratum are organized in $5$  invariant multiplets of $\mathfrak{e}_2$. Note that, for reasons of clearness, the relations generated by $\mathfrak{A}^{(0,0)}_2 $ and ${\mathfrak{A}^{(0,0)}_2}^*=\mathfrak{\bar{A}}^{(0,0)}_{-2} $, the latter ones being labeled with barred letters, are listed separately  even though they belong to the same multiplet. The same holds for  ${\mathfrak{C}^{(0,0)}_2}$ and ${\mathfrak{C}^{(0,0)}_2}^*=\mathfrak{\bar{C}}^{(0,0)}_{-2} $.  By points at the ends of the multiplets it is indicated that one can produce further relations by the action of $W_{+1}$ and $W_{-1}$ which are, however, linearly dependent on the ones already listed.

\begin{alignat*}{3}
&\mathfrak{A}&\quad = \quad&\{\R{A}{0}{0}{2}, \R{A}{1}{0}{3},\R{A}{2}{0}{4}, \R{A}{0}{1}{1},
\R{A}{1}{1}{2}, \R{A}{2}{1}{3}, \R{A}{0}{2}{0}, \R{A}{1}{2}{1},\\
&& &\R{A}{2}{2}{2}, \R{A}{0}{3}{-1},
 \R{A}{1}{3}{0}, \R{A}{2}{3}{1},
\R{A}{0}{4}{-2}, \R{A}{1}{4}{-1}, \R{A}{2}{4}{0}, \R{A}{0}{5}{-3}\}
\\
\\
&\mathfrak{\bar{A}}&\quad = \quad&\{\R{\bar{A}}{0}{0}{-2}, \R{\bar{A}}{0}{1}{-3},\R{\bar{A}}{0}{2}{-4}, \R{\bar{A}}{1}{0}{-1},\R{\bar{A}}{1}{1}{-2},\R{\bar{A}}{1}{2}{-3},
\R{\bar{A}}{2}{0}{0},\R{\bar{A}}{2}{1}{-1},\\ 
&& & \R{\bar{A}}{2}{2}{-2},\R{\bar{A}}{3}{0}{1},\R{\bar{A}}{3}{1}{0}, \R{\bar{A}}{3}{2}{-1},
\R{\bar{A}}{4}{0}{2}, \R{\bar{A}}{4}{1}{1},\quad.\quad , \R{\bar{A}}{5}{0}{3}\}\\
\\
&\mathfrak{B}&\quad = \quad&\{\R{B}{0}{0}{0}, \R{B}{1}{0}{1},\R{B}{2}{0}{2}, \R{B}{0}{1}{-1},
\R{B}{1}{1}{0}, \R{B}{2}{1}{1}, \R{B}{0}{2}{-2}, \R{B}{1}{2}{-1},\dots\}\\
\\
&\mathfrak{C}&\quad = \quad&\{\R{C}{0}{0}{2}, \R{C}{1}{0}{3},\R{C}{0}{1}{1},
 \R{C}{1}{1}{2},\dots\}\\
\\
&\mathfrak{\bar{C}}&\quad = \quad&\{\R{\bar{C}}{0}{0}{-2},\R{\bar{C}}{0}{1}{-3},\R{\bar{C}}{1}{0}{-1},
 \R{\bar{C}}{1}{1}{-2},\dots\}\\
\\
&\mathfrak{D}&\quad = \quad&\{\R{D}{0}{0}{0}, \R{D}{1}{0}{1},\R{D}{0}{1}{-1},\dots\}\\
\\
&\mathfrak{E}&\quad = \quad&\{\R{E}{0}{0}{0},\dots\}. 
\end{alignat*}
Of course, this is not the only way to organize the 51 constituent relations but it is impossible to organize them without overlapping multiplets or with less than $5$ multiplets.
\section{Summary and Conclusion}
The  algebra of classical observables of the closed bosonic string is analyzed for the case $P^2=0,P\neq0$. An algebraically independent set of invariants, called standard invariants, is constructed generating via multiplication the algebra of invariants. This construction is done analogously to the construction of the standard invariants of the massive string due to the choice of an adapted basis in Minkowski space. The algebra of invariants for the massless string carries a representation of the stabilizer group $E_2$ of light-like $P$ in its standard form. It turns out that, unlike in the massive case, the linear span of standard invariants of the first stratum (and, accordingly, of strata of higher degree) does not provide a representation space of $E_2$ on its own, in contrast to  the vector space of the first stratum as a whole, including the products of the elements of the zeroth stratum. The standard invariants of the first and second stratum can only be arranged in multiplets of $E_2$ which show an improper overlap.  This fact complicates the task of presenting  the algebra as has been illustrated here for the first two strata. The results obtained demonstrate that the massless string cannot just be treated as a limit of the massive string. 
\\ \\ \\
\emph{Acknowledgment.} I would like to thank Professor K. Pohlmeyer who supervised my diploma thesis on the subject investigated here. I am also thankful for his critical comments on this manuscript.
\\ \\ \\

\section*{Appendix:~The Standard Invariants of the $2$.~Stratum in Multiplets of $\boldsymbol{\mathfrak{e}_2}$}

\begin{alignat*}{2}
&V &=\quad&\{\In{V}{0}{0}{1},\In{V}{1}{0}{2},\In{V}{2}{0}{3},\In{V}{0}{1}{0},\In{V}{1}{1}{1},
\In{V}{2}{1}{2},\In{V}{0}{2}{-1},\In{V}{1}{2}{0},\In{V}{2}{2}{1},\dots\}\\
\\
&\bar{V} &=\quad& \{\In{\bar{V}}{0}{0}{-1},\In{\bar{V}}{0}{1}{-2},\In{\bar{V}}{0}{2}{-3},\In{\bar{V}}{1}{0}{0},\In{\bar{V}}{1}{1}{-1},\In{\bar{V}}{1}{2}{-2},\In{\bar{V}}{2}{0}{1},
 \In{\bar{V}}{2}{1}{0},
 \In{\bar{V}}{2}{2}{-1},\dots\}\\\\
&&\mbox{with }& \In{V}{0}{0}{1} =
\{\In{T}{0}{0}{0},\In{T}{1}{0}{1}\} = -4\Z^{(3)}_{\circ\circ
  -\circ++}-4\Z^{(3)}_{\circ\circ +\circ-+}
-\frac{2}{3}\Z^{(2)}_{\circ+ \circ\circ\times}\\ 
\end{alignat*}
where $\In{\bar{V}}{0}{0}{-1}$ is given by ${V^{(0,0)}_1}^*$ and  $\In{V}{p}{q}{m}$ and $\In{\bar{V}}{p}{q}{n}$ are obtained by repeated action of $W_{+1}$ (p-fold) and $W_{-1}$ (q-fold) on $\In{V}{0}{0}{1}$ and $\In{\bar{V}}{0}{0}{-1}$, respectively. The same notation is used for the following multiplets.
\\ 
\begin{alignat*}{3}
&B&=\quad&\{\In{B}{0}{0}{0},\In{B}{1}{0}{1},\In{B}{2}{0}{2},\In{B}{0}{1}{-1},\In{B}{1}{1}{0},
 \In{B}{2}{1}{1},\In{B}{0}{2}{-2},\In{B}{1}{2}{-1},\In{B}{2}{2}{0},\dots\}\\\\
&&\mbox{with }& 
  \In{B}{0}{0}{0}=
 \{\In{A}{0}{0}{2},\In{\bar{A}}{0}{0}{-2}\}=
 \frac{8}{3}\Z^{(2)}_{\circ- \circ\circ+}\\ \\
 &E&=\quad&\{\In{E}{0}{0}{1},\In{E}{1}{0}{2},\In{E}{0}{1}{0},\In{E}{1}{1}{1},
 \In{E}{0}{2}{-1},\In{E}{1}{2}{0}\}\\ \\
&\bar{E}&=\quad&\{\In{\bar{E}}{0}{0}{-1},\In{\bar{E}}{0}{1}{-2},\In{\bar{E}}{1}{0}{0},\In{\bar{E}}{1}{1}{-1},\In{\bar{E}}{2}{0}{1}, \dots\}\\ \\
&&\mbox{with }&
 \In{E}{0}{0}{1}=\{\In{A}{0}{1}{1},\In{T}{0}{0}{0}\}=
 -\frac{4}{3}\Z^{(2)}_{\circ+ \circ\circ\times}
 -8\Z^{(3)}_{\circ\circ +-\circ+}-24\Z^{(3)}_{\circ\circ -\circ++}\\
&&\mbox{and }&
 \In{\bar{E}}{0}{0}{-1}={\In{E}{0}{0}{1}}^*\\
\end{alignat*}
\begin{alignat*}{2}
&H &=\quad& \{\In{H}{0}{0}{3},\dots\}\\\\
&\bar{H} &=\quad& \{\In{\bar{H}}{0}{0}{-3},\dots\}\\
\\
&&\mbox{with }& 
 \In{H}{0}{0}{3}=\{\In{A}{0}{0}{2},\In{A}{0}{1}{1}\}=
 32\Z^{(3)}_{\circ\circ +\circ++}\\
&&\mbox{and }&\In{\bar{H}}{0}{0}{-3}={\In{H}{0}{0}{3}}^*.\\
\end{alignat*}

\end{document}